\documentclass[12pt]{article}
\usepackage{graphicx}

\begin{document}

\title{Direct derivation of the Peres-Horodecki criterion for the two-qubit states
from
the Hill-Wootters formula for the entanglement of formation}

\author{Hiroo Azuma${}^{1,}$\thanks{Email: hiroo.azuma@m3.dion.ne.jp}
\ \ 
and
\ \ 
Masashi Ban${}^{2,}$\thanks{Email: m.ban@phys.ocha.ac.jp}
\\
\\
{\small ${}^{1}$Information and Mathematical Science Laboratory Inc.,}\\
{\small Meikei Bldg., 1-5-21 Ohtsuka, Bunkyo-ku, Tokyo 112-0012, Japan}
\\
{\small ${}^{2}$Graduate School of Humanities and Sciences, Ochanomizu University,}\\
{\small 2-1-1 Ohtsuka, Bunkyo-ku, Tokyo 112-8610, Japan}
}

\date{\today}

\maketitle

\begin{abstract}
In this paper, we show a direct method
of deriving the Peres-Horodecki criterion for the two-qubit states
from the Hill-Wootters formula for the entanglement of formation.
Although the Peres-Horodecki criterion and the Hill-Wootters formula are
established results in the field of quantum information theory,
they are proved independently and
connections between them are not discussed precisely.
In this paper, we clarify these connections.
First, we replace the original Peres-Horodecki criterion
with its another equivalent statement
found by Augusiak {\it et al}.
Second, we obtain an analytical form of the concurrence
of an arbitrary two-qubit state $\rho$,
using Ferrari's method to solve a quartic equation for eigenvalues
$\rho\tilde{\rho}$.
Finally, with the above preparations,
we accomplish the direct derivation of the Peres-Horodecki criterion
from the Hill-Wootters formula.
\end{abstract}

\bigskip


\bigskip

\section{Introduction}\label{section-introduction}
Since Einstein, Podolsky and Rosen pointed out the importance
of entanglement that appears in the quantum state of the bipartite system,
many researchers have been investigating the nature of the entanglement
\cite{Einstein1935}.
Because it is understood that the entanglement can be a resource of
quantum information processing,
both classifying and quantifying the entanglement have become topics which attract wide audience
in the field of quantum information
\cite{Bennett1993}.

To investigate the entanglement theoretically,
we can take two different approaches.
One is the qualitative analysis,
such as concentrating on how to distinguish
entangled (inseparable) states of the bipartite system from separable states.
The other is the quantitative analysis,
such as finding an appropriate measure of entanglement
to represent the entanglement as an amount numerically.

The Peres-Horodecki criterion belongs in the former approach
\cite{Peres1996, Horodeckis1996}.
It gives the necessary and sufficient conditions for
separability of mixed states of two-qubit systems and qubit-qutrit systems.
At first, Peres obtains this criterion as a conjecture and proves
that it is a necessary condition of separability
\cite{Peres1996}.
Next, Horodecki {\it et al}. prove that
this criterion is sufficient condition of separability
for two-qubit systems and qubit-qutrit systems
\cite{Horodeckis1996}.
To follow the proof of reference~\cite{Horodeckis1996},
we have to utilize some results of pure mathematics,
for example, the Hahn-Banach theorem in the functional analysis
and Str{\o}mer and Woronowicz's results about positive maps
\cite{Aubin1999, Stromer1963, Woronowicz1976}.
Thus, in spite of its simple form,
the derivation of the Peres-Horodecki criterion is not accessible to
physicists.

The Hill-Wootters formula for the entanglement of formation
belongs in the latter approach
\cite{Hill1997, Wootters1998}.
It gives an explicit formula of the entanglement of formation of the two-qubit mixed states.
(The entanglement of formation of a mixed state is defined
as the minimum average entanglement of an ensemble of pure states
that represents the original mixed state.)
To obtain this formula,
we use general properties of entropy,
so that the derivation of the Hill-Wootters formula is accessible to physicists.

As mentioned above,
the Peres-Horodecki criterion and the Hill-Wootters formula
are derived independently with each other.
However,
deriving the Peres-Horodecki criterion
from the Hill-Wootters formula in a direct manner
has to be possible in principle,
so that there must be connections between them.
This is the motivation of this paper.

In this paper,
we show a direct method
of deriving the Peres-Horodecki criterion for the two-qubit mixed states from
the Hill-Wootters formula.
First, we replace the original Peres-Horodecki criterion
with the following equivalent statement obtained by
Augusiak {\it et al}.
\cite{Augusiak2008, Horodecki2009}:
an arbitrary two-qubit mixed state is inseparable
if and only if
the determinant of the partial transpose of its density matrix is less than zero.
Second,
we obtain an analytical form of the concurrence of an arbitrary two-qubit state
$\rho$,
using Ferrari's method to solve a quartic equation for eigenvalues $\rho\tilde{\rho}$.
Finally, with the above preparations,
we accomplish the direct derivation of the Peres-Horodecki criterion from
the Hill-Wootters formula.

In the middle of the above derivation,
we show an alternative method of obtaining
Augusiak {\it et al}.'s results
\cite{Augusiak2008, Horodecki2009}.
Using the so-called Lewenstein-Sanpera decomposition and Weyl's inequality
\cite{Lewenstein1998, Bhatia1996},
we show that an arbitrary two-qubit mixed state is inseparable
if and only if the partial transpose of its density matrix has three positive
eigenvalues and one negative eigenvalue.
Thus, it never has zero eigenvalues.
In contrast, the partial transpose of a density matrix of a separable two-qubit state is
positive-semidefinite.
Hence, we reach at Augusiak {\it et al}.'s two-qubit separability condition.

Here, we refer to previous works
that relate to our results.
Sanpera {\it et al}. show
that the partial transpose of a density matrix of
an arbitrary inseparable two-qubit state
has at most one negative eigenvalue
\cite{Sanpera1998}.
Hayden obtains similar results
\cite{Hayden2000}.
Vidal and Werner propose negativity,
which is a sum of the absolute values of all negative
eigenvalues of the partially transposed density matrix,
as a measure of entanglement
\cite{Vidal2002}.
Augusiak {\it et al}. show that an arbitrary two-qubit mixed state is inseparable
if and only if the determinant of the partial transpose
of its density matrix is
less than zero
\cite{Augusiak2008, Horodecki2009}.
(Augusiak {\it et al}. pointed out
that the partial transpose of an inseparable two-qubit density matrix never has zero eigenvalues.)

Recently,
two-qubit X-states,
whose non-zero elements of the density matrix are in an ``X" formation,
are studied eagerly in relation to the investigation of the entanglement sudden death
\cite{Yu2007, Huang2007, Ali2009}.
In these works,
it is confirmed that
the negativity is essentially equivalent to the concurrence
on condition that the density matrix is in the two-qubit X-state.
Because the Werner states are a subclass of the X-states,
this result is interesting from the viewpoint of the quantum information theory.

This paper is organized as follows.
In the remains of this section,
we give brief reviews of
the Peres-Horodecki criterion
and
the Hill-Wootters formula.
In section~\ref{section-eigenvalue-flip-Peres-Horodecki},
we show that an arbitrary two-qubit mixed state is inseparable
if and only if the partial transpose of its density matrix
has three positive eigenvalues and one negative eigenvalue.
In section~\ref{section-general-case-Hill-Wootters-Peres-Horodecki},
we show a direct method of deriving the Peres-Horodecki criterion
from the Hill-Wootters formula for general two-qubit mixed states.
In section~\ref{section-convex-combination-separable-inseparable},
we consider an example of a convex combination of a separable pure state and an inseparable pure state.
Assuming that the two-qubit system is in this state,
we show that we can derive the Peres-Horodecki criterion from the Hill-Wootters formula
without difficulty.
In section~\ref{section-discussions},
we give brief discussions.
In appendix~\ref{section-appendix},
we show some results
of calculations obtained in section~\ref{section-general-case-Hill-Wootters-Peres-Horodecki}.

In this paper,
we define the separability of the two-qubit system AB as follows:
if the density matrix $\rho_{\mbox{\scriptsize AB}}$ can be
written as a convex combination of product states,
\begin{equation}
\rho_{\mbox{\scriptsize AB}}
=
\sum_{i}p_{i}\rho_{\mbox{\scriptsize A},i}\otimes\rho_{\mbox{\scriptsize B},i},
\end{equation}
where
$p_{i}\geq 0$,
$\mbox{tr}\rho_{\mbox{\scriptsize A},i}=1$,
$\mbox{tr}\rho_{\mbox{\scriptsize B},i}=1$ $\forall i$
and $\sum_{i}p_{i}=1$,
then $\rho_{\mbox{\scriptsize AB}}$ is separable.
If and only if $\rho_{\mbox{\scriptsize AB}}$ is not separable,
it is inseparable
\cite{Werner1989}.

Here, we give exact descriptions of the
Peres-Horodecki criterion
and the Hill-Wootters formula.

The Peres-Horodecki criterion for the two-qubit states
is described as follows.
We let $\rho_{\mbox{\scriptsize AB}}$ be a density matrix
of an arbitrary mixed state
of the two-qubit system AB.
$\rho_{\mbox{\scriptsize AB}}^{\mbox{\scriptsize PT}}$ denotes
the partial transpose of $\rho_{\mbox{\scriptsize AB}}$
with respect to the qubit B.
The necessary and sufficient condition for separability of $\rho_{\mbox{\scriptsize AB}}$
is the positivity of $\rho_{\mbox{\scriptsize AB}}^{\mbox{\scriptsize PT}}$.

The Hill-Wootters formula for the entanglement of formation of the two-qubit states
is described as follows.
We let $\rho_{\mbox{\scriptsize AB}}$ be a density matrix
of an arbitrary mixed state
of the two-qubit system AB.
We write matrix elements of $\rho_{\mbox{\scriptsize AB}}$
in the fixed basis\\
$\{
|0\rangle_{\mbox{\scriptsize A}}|0\rangle_{\mbox{\scriptsize B}},
|0\rangle_{\mbox{\scriptsize A}}|1\rangle_{\mbox{\scriptsize B}},
|1\rangle_{\mbox{\scriptsize A}}|0\rangle_{\mbox{\scriptsize B}},
|1\rangle_{\mbox{\scriptsize A}}|1\rangle_{\mbox{\scriptsize B}}
\}$.
We describe $\rho_{\mbox{\scriptsize AB}}^{*}$ as the complex conjugate
of $\rho_{\mbox{\scriptsize AB}}$
in this fixed basis.
Moreover, we define a new matrix
$\tilde{\rho}_{\mbox{\scriptsize AB}}
=(\sigma_{y,\mbox{\scriptsize A}}\otimes\sigma_{y,\mbox{\scriptsize B}})
\rho_{\mbox{\scriptsize AB}}^{*}
(\sigma_{y,\mbox{\scriptsize A}}\otimes\sigma_{y,\mbox{\scriptsize B}})$.
Next,
we write four eigenvalues of $\rho_{\mbox{\scriptsize AB}}\tilde{\rho}_{\mbox{\scriptsize AB}}$
as
$\lambda_{1}$, $\lambda_{2}$, $\lambda_{3}$ and $\lambda_{4}$
in decreasing order.
Because of the definition of $\tilde{\rho}_{\mbox{\scriptsize AB}}$,
we have the relation
$\lambda_{i}\geq 0$ $\forall i$
for any $\rho_{\mbox{\scriptsize AB}}$.
Hence, we can expect
$\lambda_{1}\geq\lambda_{2}\geq\lambda_{3}\geq\lambda_{4}\geq 0$.

After these preparations,
we define the concurrence,
$C(\rho_{\mbox{\scriptsize AB}})
=\mbox{max}
\{0,
\sqrt{\lambda_{1}}-\sqrt{\lambda_{2}}-\sqrt{\lambda_{3}}-\sqrt{\lambda_{4}}\}$.
If $\rho_{\mbox{\scriptsize AB}}$ represents a pure state,
we obtain
$C(|\psi\rangle_{\mbox{\scriptsize AB}})
=|_{\mbox{\scriptsize AB}}\langle\psi|\tilde{\psi}\rangle_{\mbox{\scriptsize AB}}|$.
Furthermore,
we can obtain the relation
$0\leq C(\rho_{\mbox{\scriptsize AB}})\leq 1$
for an arbitrary mixed state $\rho_{\mbox{\scriptsize AB}}$.
The entanglement of formation is defined
as $E(\rho_{\mbox{\scriptsize AB}})={\cal E}(C(\rho_{\mbox{\scriptsize AB}}))$,
where
\begin{eqnarray}
{\cal E}(C)
&=&
-\frac{1+\sqrt{1-C^{2}}}{2}
\log_{2}
\frac{1+\sqrt{1-C^{2}}}{2} \nonumber \\
&&
-\frac{1-\sqrt{1-C^{2}}}{2}
\log_{2}
\frac{1-\sqrt{1-C^{2}}}{2}.
\end{eqnarray}
The function ${\cal E}(C)$ increases monotonically
and varies form zero to unity as $C$ goes from zero to unity.
Thus, if and only if $C(\rho_{\mbox{\scriptsize AB}})=0$,
$\rho_{\mbox{\scriptsize AB}}$ is separable.

\section{The number of negative eigenvalues of the partially transposed inseparable density matrix}
\label{section-eigenvalue-flip-Peres-Horodecki}
In this section,
we show that the partial transpose
of the density matrix of the inseparable mixed state for the two-qubit system
always has
three positive eigenvalues and one negative eigenvalue,
so that it never has zero eigenvalues.
Then,
we obtain another expression which is equivalent to
the Peres-Horodecki criterion for the two-qubit states
as follows:
the two-qubit mixed state is inseparable
if and only if the determinant of the partial transpose of its density matrix
is negative.
(As mentioned in section~\ref{section-introduction},
we show a simple method of deriving Augusiak {\it et al}.'s results
\cite{Augusiak2008, Horodecki2009}.)

First of all,
we consider the separability and inseparability of an arbitrary two-qubit pure state
$|\psi\rangle_{\mbox{\scriptsize AB}}$.
We can describe $|\psi\rangle_{\mbox{\scriptsize AB}}$
as the following four-element ket vector without loosing generality,
\begin{equation}
|\psi\rangle_{\mbox{\scriptsize AB}}
=
\left(
\begin{array}{c}
a \\
b \exp(i\theta_{1}) \\
c \exp(i\theta_{2}) \\
\sqrt{1-a^{2}-b^{2}-c^{2}} \exp(i\theta_{3})
\end{array}
\right),
\end{equation}
where the basis is given by
$\{
|0\rangle_{\mbox{\scriptsize A}}|0\rangle_{\mbox{\scriptsize B}},
|0\rangle_{\mbox{\scriptsize A}}|1\rangle_{\mbox{\scriptsize B}},
|1\rangle_{\mbox{\scriptsize A}}|0\rangle_{\mbox{\scriptsize B}},
|1\rangle_{\mbox{\scriptsize A}}|1\rangle_{\mbox{\scriptsize B}}
\}$
and we assume
$a\geq 0$, $b\geq 0$, $c\geq 0$, $1-a^{2}-b^{2}-c^{2}\geq 0$,
$0 \leq \theta_{i}<2\pi$ for $i\in\{1,2,3\}$.

By tracing out the freedom of the qubit B,
we obtain the reduced density matrix of the qubit A as
\begin{eqnarray}
\rho_{\mbox{\scriptsize A}}
&=&
\mbox{tr}_{\mbox{\scriptsize B}}
(|\psi\rangle_{\mbox{\scriptsize AB}}{}_{\mbox{\scriptsize AB}}\langle \psi|) \nonumber \\
&=&
\left(
\begin{array}{cc}
\rho_{\mbox{\scriptsize A},00} & \rho_{\mbox{\scriptsize A},01} \\
\rho_{\mbox{\scriptsize A},01}^{*} & 1-\rho_{\mbox{\scriptsize A},00}
\end{array}
\right),
\end{eqnarray}
where
\begin{eqnarray}
\rho_{\mbox{\scriptsize A},00}
&=&
a^{2}+b^{2}, \nonumber \\
\rho_{\mbox{\scriptsize A},01}
&=&
ac\exp(-i\theta_{2})+b\sqrt{1-a^{2}-b^{2}-c^{2}}\exp[i(\theta_{1}-\theta_{3})].
\end{eqnarray}
Then, the eigenvalues of $\rho_{\mbox{\scriptsize A}}$ are given by
\begin{equation}
\lambda_{\pm}=\frac{1}{2}(1\pm\sqrt{1-C^{2}}),
\label{eigenvalues-rho-A}
\end{equation}
where
\begin{eqnarray}
C
&=&
2
[
a^{2}(1-a^{2}-b^{2}-c^{2})+b^{2}c^{2} \nonumber \\
&&\quad
-2abc\sqrt{1-a^{2}-b^{2}-c^{2}}\cos(\theta_{1}+\theta_{2}-\theta_{3})
]^{1/2}.
\label{concurrence-pure-state-0}
\end{eqnarray}
The quantity $C$ in equation~(\ref{concurrence-pure-state-0})
is equal to the concurrence of $|\psi\rangle_{\mbox{\scriptsize AB}}$
given by the Hill-Wootters formula.
Because $\rho_{\mbox{\scriptsize A}}$ is positive-semidefinite
and $\mbox{tr}\rho_{\mbox{\scriptsize A}}=1$,
we obtain $0\leq\lambda_{\pm}\leq 1$.
Thus, from equation~(\ref{eigenvalues-rho-A}),
we obtain $0\leq C\leq 1$.
If $C=0$, $|\psi\rangle_{\mbox{\scriptsize AB}}$ is separable.
In contrast, if $C>0$, $|\psi\rangle_{\mbox{\scriptsize AB}}$ is inseparable.

On the other hand,
writing the density matrix as
$\rho_{\mbox{\scriptsize AB}}
=|\psi\rangle_{\mbox{\scriptsize AB}}{}_{\mbox{\scriptsize AB}}\langle\psi |$,
we obtain four eigenvalues of
$\rho_{\mbox{\scriptsize AB}}^{\mbox{\scriptsize PT}}$
as follows:
\begin{eqnarray}
\eta_{1}&=&(1/2)C, \nonumber \\
\eta_{2}&=&-(1/2)C, \nonumber \\
\eta_{3}&=&(1/2)(1+\sqrt{1-C^{2}}), \nonumber \\
\eta_{4}&=&(1/2)(1-\sqrt{1-C^{2}}).
\end{eqnarray}

From the above results
and the relation $0\leq C\leq 1$,
we can conclude that
$\rho_{\mbox{\scriptsize AB}}^{\mbox{\scriptsize PT}}$
is positive-semidefinite
if and only if $C(|\psi\rangle_{\mbox{\scriptsize AB}})=0$.
At the same time,
we find that
$\rho_{\mbox{\scriptsize AB}}^{\mbox{\scriptsize PT}}$
has three positive eigenvalues and one negative eigenvalue
if and only if $C(|\psi\rangle_{\mbox{\scriptsize AB}})\neq 0$.

Hence,
we obtain the following fact.
If and only if an arbitrary pure state of the two-qubit system
$|\psi\rangle_{\mbox{\scriptsize AB}}$
is inseparable,
$\rho^{\mbox{\scriptsize PT}}_{\mbox{\scriptsize AB}}
=(|\psi\rangle_{\mbox{\scriptsize AB}}{}_{\mbox{\scriptsize AB}}\langle\psi |)^{\mbox{\scriptsize PT}}$
has three positive eigenvalues and one negative eigenvalue.

Next, we use the following result obtained by Lewenstein and Sanpera
\cite{Lewenstein1998}.
An arbitrary two-qubit mixed state $\rho$ has a decomposition
in the form,
\begin{equation}
\rho=p\rho_{s}+(1-p)\rho_{e},
\label{Lewenstein-Sanpera-decomposition}
\end{equation}
where
$0\leq p \leq 1$,
$\rho_{s}$ is a normalized separable mixed state
and
$\rho_{e}=|\psi\rangle\langle\psi |$ is a normalized inseparable pure state.
[Here, we pay attention to the following fact.
The decomposition of equation~(\ref{Lewenstein-Sanpera-decomposition})
is obtained with neither the Peres-Horodecki criterion
nor the Hill-Wootters formula.
Thus, the so-called Lewenstein-Sanpera decomposition is derived independently
from the Peres-Horodecki criterion and the Hill-Wootters formula.]

Let us take the partial transpose of equation~(\ref{Lewenstein-Sanpera-decomposition}),
\begin{equation}
\rho^{\mbox{\scriptsize PT}}
=
p\rho_{s}^{\mbox{\scriptsize PT}}
+
(1-p)\rho_{e}^{\mbox{\scriptsize PT}}.
\label{Lewenstein-Sanpera-decomposition-PT}
\end{equation}
According to the Peres-Horodecki criterion,
all of the four eigenvalues of $\rho_{s}^{\mbox{\scriptsize PT}}$
are equal to or larger than zero.
Thus, we write the eigenvalues of $p\rho_{s}^{\mbox{\scriptsize PT}}$
as
\begin{equation}
\eta_{1}^{\downarrow}(p\rho_{s}^{\mbox{\scriptsize PT}})
\geq
\eta_{2}^{\downarrow}(p\rho_{s}^{\mbox{\scriptsize PT}})
\geq
\eta_{3}^{\downarrow}(p\rho_{s}^{\mbox{\scriptsize PT}})
\geq
\eta_{4}^{\downarrow}(p\rho_{s}^{\mbox{\scriptsize PT}})
\geq
0,
\label{eigenvalues-separable-density-matrix}
\end{equation}
in decreasing order.
In contrast, from the fact obtained before,
$\rho_{e}^{\mbox{\scriptsize PT}}=(|\psi\rangle\langle\psi |)^{\mbox{\scriptsize PT}}$
has three positive eigenvalues and one negative eigenvalue.
Thus,
we write the eigenvalues of $(1-p)\rho_{e}^{\mbox{\scriptsize PT}}$
as
\begin{eqnarray}
&&
\eta_{1}^{\downarrow}((1-p)\rho_{e}^{\mbox{\scriptsize PT}})
\geq
\eta_{2}^{\downarrow}((1-p)\rho_{e}^{\mbox{\scriptsize PT}})
\geq
\eta_{3}^{\downarrow}((1-p)\rho_{e}^{\mbox{\scriptsize PT}}) \nonumber \\
&>&
0
>
\eta_{4}^{\downarrow}((1-p)\rho_{e}^{\mbox{\scriptsize PT}})
\quad\quad
\mbox{for $p\neq 1$},
\label{eigenvalues-entangled-density-matrix}
\end{eqnarray}
in decreasing order.

Here, we use Weyl's inequality
given by the following form
\cite{Bhatia1996}.
We let $X$ and $Y$ be arbitrary $n\times n$ Hermitian matrices.
Then,
\begin{eqnarray}
\lambda_{j}^{\downarrow}(X+Y)
&\leq &
\lambda_{i}^{\downarrow}(X)
+
\lambda_{j-i+1}^{\downarrow}(Y)
\quad\quad
\mbox{for $i\leq j$}, \nonumber \\
\lambda_{j}^{\downarrow}(X+Y)
&\geq &
\lambda_{i}^{\downarrow}(X)
+
\lambda_{j-i+n}^{\downarrow}(Y)
\quad\quad
\mbox{for $i\geq j$}.
\label{Weyl-inequality}
\end{eqnarray}
Thus,
from equations~(\ref{eigenvalues-separable-density-matrix}),
(\ref{eigenvalues-entangled-density-matrix})
and
(\ref{Weyl-inequality}),
we obtain
\begin{equation}
\eta_{3}^{\downarrow}(\rho^{\mbox{\scriptsize PT}})
\geq
\eta_{4}^{\downarrow}(p\rho_{s}^{\mbox{\scriptsize PT}})
+
\eta_{3}^{\downarrow}((1-p)\rho_{e}^{\mbox{\scriptsize PT}})
>
0
\quad
\mbox{for $p\neq 1$}.
\end{equation}
Hence,
if $\rho$ is inseparable,
$\rho^{\mbox{\scriptsize PT}}$ has three positive eigenvalues.

On the other hand,
the Peres-Horodecki criterion tells us that
$\rho^{\mbox{\scriptsize PT}}$
has negative eigenvalues
if and only if $\rho$ is inseparable.
Therefore,
if the two-qubit mixed state $\rho$ is inseparable,
its partial transpose has only one negative eigenvalue.
Moreover, the other eigenvalues are positive and
$\rho^{\mbox{\scriptsize PT}}$ never has zero eigenvalues,
when $\rho$ is inseparable.

From the above discussions,
we obtain another expression
which is equivalent to the Peres-Horodecki criterion for two-qubit mixed states
as follows:
a two-qubit mixed state is inseparable
if and only if the determinant of the partial transpose
of its density matrix is negative.
Because we obtain this statement,
we change our purpose of this paper into new one.
From next section,
we replace the original Peres-Horodecki criterion with this new expression
and
we try to derive this new expression
from the Hill-Wootters formula
in a direct manner.

\section{Direct derivation of the Peres-Horodecki criterion
from the Hill-Wootters formula
for the general\\
two-qubit mixed states}\label{section-general-case-Hill-Wootters-Peres-Horodecki}
In this section,
we show the direct derivation of the Peres-Horodecki criterion
from the Hill-Wootters formula
for general two-qubit mixed states.

First of all, we give a density matrix of an arbitrary two-qubit mixed state
as
\begin{equation}
\rho_{\mbox{\scriptsize AB}}
=
\left(
\begin{array}{cccc}
r_{0} & u_{0}e^{i\theta_{0}} & u_{1}e^{i\theta_{1}} & u_{2}e^{i\theta_{2}} \\
u_{0}e^{-i\theta_{0}} & r_{1} & u_{3}e^{i\theta_{3}} & u_{4}e^{i\theta_{4}} \\
u_{1}e^{-i\theta_{1}} & u_{3}e^{-i\theta_{3}} & r_{2} & u_{5}e^{i\theta_{5}} \\
u_{2}e^{-i\theta_{2}} & u_{4}e^{-i\theta_{4}} & u_{5}e^{-i\theta_{5}} & 1-r_{0}-r_{1}-r_{2}
\end{array}
\right),
\label{general-2-qubit-density-matrix-0}
\end{equation}
where
the basis is given by
$\{|0\rangle_{\mbox{\scriptsize A}}|0\rangle_{\mbox{\scriptsize B}},
|0\rangle_{\mbox{\scriptsize A}}|1\rangle_{\mbox{\scriptsize B}},
|1\rangle_{\mbox{\scriptsize A}}|0\rangle_{\mbox{\scriptsize B}},
|1\rangle_{\mbox{\scriptsize A}}|1\rangle_{\mbox{\scriptsize B}}\}$,
$r_{i}\geq 0$ for $i\in\{0,1,2\}$, $1-r_{0}-r_{1}-r_{2}\geq 0$,
$u_{j}\geq 0$ for $j\in\{0,1,...,5\}$ and
$0\leq\theta_{k}<2\pi$ for $k\in\{0,1,...,5\}$.

Next, we remove some real parameters,
which describe the local freedom of each qubit,
from the expression of equation~(\ref{general-2-qubit-density-matrix-0}).
We think about the following submatrix,
which is obtained by setting a basis vector of the qubit A on $|0\rangle_{\mbox{\scriptsize A}}$,
\begin{equation}
\left(
\begin{array}{cc}
r_{0} & u_{0}\exp(i\theta_{0}) \\
u_{0}\exp(-i\theta_{0}) & r_{1}
\end{array}
\right).
\end{equation}
We can always diagonalize this submatrix
by applying a certain SU(2) rotation to the basis of the qubit B,
$\{|0\rangle_{\mbox{\scriptsize B}},|1\rangle_{\mbox{\scriptsize B}}\}$.
In a similar way,
we think about the following matrix,
which is obtained by setting a basis vector of the qubit B on $|0\rangle_{\mbox{\scriptsize B}}$,
\begin{equation}
\left(
\begin{array}{cc}
r_{0} & u_{1}\exp(i\theta_{1}) \\
u_{1}\exp(-i\theta_{1}) & r_{2}
\end{array}
\right).
\end{equation}
We can always diagonalize this submatrix
by applying a certain SU(2) rotation to the basis of the qubit A,
$\{|0\rangle_{\mbox{\scriptsize A}},|1\rangle_{\mbox{\scriptsize A}}\}$,
as well.
We can take these two local SU(2) rotations independently with each other,
and these transformations never cause effects on the entanglement between the qubits A and B.
Thus, these local transformations never give rise to effects on both
the Peres-Horodecki criterion and the Hill-Wootters formula.

From these discussions,
we can replace the form of $\rho_{\mbox{\scriptsize AB}}$ written
in equation~(\ref{general-2-qubit-density-matrix-0})
with the following form,
\begin{equation}
\rho_{\mbox{\scriptsize AB}}
=
\left(
\begin{array}{cccc}
r_{0} & 0 & 0 & u_{2}e^{i\theta_{2}} \\
0 & r_{1} & u_{3}e^{i\theta_{3}} & u_{4}e^{i\theta_{4}} \\
0 & u_{3}e^{-i\theta_{3}} & r_{2} & u_{5}e^{i\theta_{5}} \\
u_{2}e^{-i\theta_{2}} & u_{4}e^{-i\theta_{4}} & u_{5}e^{-i\theta_{5}} & 1-r_{0}-r_{1}-r_{2}
\end{array}
\right),
\label{general-2-qubit-density-matrix-1}
\end{equation}
where
$\{r_{i}\}$, $\{u_{i}\}$, $\{\theta_{i}\}$
given in equation~(\ref{general-2-qubit-density-matrix-1})
do not need to be equal to
$\{r_{i}\}$, $\{u_{i}\}$, $\{\theta_{i}\}$
given in equation~(\ref{general-2-qubit-density-matrix-0}).

Moreover, we apply the following local SU(2) transformation
to $\rho_{\mbox{\scriptsize AB}}$ defined in equation~(\ref{general-2-qubit-density-matrix-1}),
\begin{equation}
\left\{
\begin{array}{lll}
|0\rangle_{\mbox{\scriptsize A}} & \to & |0\rangle_{\mbox{\scriptsize A}}, \\
|1\rangle_{\mbox{\scriptsize A}} & \to & \exp(i\theta_{3})|1\rangle_{\mbox{\scriptsize A}}.
\end{array}
\right.
\end{equation}
By this transformation,
we can change complex numbers of the $|01\rangle\langle 10|$- and $|10\rangle\langle 01|$- entries
of $\rho_{\mbox{\scriptsize AB}}$ defined in equation~(\ref{general-2-qubit-density-matrix-1})
into real numbers
as $u_{3}\exp(\pm i\theta_{3}) \to u_{3}$.
Thus,
we obtain the final form of the general $\rho_{\mbox{\scriptsize AB}}$ as
\begin{equation}
\rho_{\mbox{\scriptsize AB}}
=
\left(
\begin{array}{cccc}
r & 0 & 0 & u\exp(i\tau_{1}) \\
0 & s & v & w\exp(i\tau_{2}) \\
0 & v & t & q\exp(i\tau_{3}) \\
u\exp(-i\tau_{1}) & w\exp(-i\tau_{2}) & q\exp(-i\tau_{3}) & 1-r-s-t
\end{array}
\right),
\label{general-2-qubit-density-matrix-2}
\end{equation}
where
$r\geq 0$, $s\geq 0$, $t\geq 0$, $1-r-s-t\geq 0$,
$u\geq 0$, $v\geq 0$, $w\geq $, $q\geq 0$,
$0\leq\tau_{i}<2\pi$ for $i\in\{1,2,3\}$.

Because $\rho_{\mbox{\scriptsize AB}}$ defined in equation~(\ref{general-2-qubit-density-matrix-2})
is a density matrix,
$\rho_{\mbox{\scriptsize AB}}$ has to be positive-semidefinite.
Especially,
its determinant is always equal to or larger than zero,
so that we obtain
\begin{eqnarray}
\det(\rho_{\mbox{\scriptsize AB}})
&=&
-rsq^{2}-rtw^{2}
+[r(1-r-s-t)-u^{2}](st-v^{2}) \nonumber \\
&&
+2rvwq\cos(\tau_{2}-\tau_{3}) \nonumber \\
&\geq &
0.
\label{determinant-positive-or-zero-0}
\end{eqnarray}
This relation is used in the latter half of this section.

Here, we pay attention to the fact that $\rho_{\mbox{\scriptsize AB}}$
defined in equation~(\ref{general-2-qubit-density-matrix-2})
includes ten real parameters.
General two-qubit mixed states include fifteen real parameters
because of the degree of freedom of SU(4).
However, each local qubit has three real parameters
which come from the degree of freedom of SU(2).
Thus, to describe the entanglement of the two-qubit system,
we need nine real parameters.
In fact, Luo suggests the following form for $\rho_{\mbox{\scriptsize AB}}$
to investigate the entanglement
\cite{Luo2008},
\begin{equation}
\rho_{\mbox{\scriptsize AB}}
=
\frac{1}{4}
(I_{4,\mbox{\scriptsize AB}}
+
\sum_{i=1}^{3}a_{i}\sigma_{i,\mbox{\scriptsize A}}\otimes I_{2,\mbox{\scriptsize B}}
+
\sum_{i=1}^{3}b_{i}I_{2,\mbox{\scriptsize A}}\otimes \sigma_{i,\mbox{\scriptsize B}}
+
\sum_{i=1}^{3}c_{i}\sigma_{i,\mbox{\scriptsize A}}\otimes \sigma_{i,\mbox{\scriptsize B}}).
\label{general-2-qubit-density-matrix-3}
\end{equation}
However, we dare to choose equation~(\ref{general-2-qubit-density-matrix-2})
rather than equation~(\ref{general-2-qubit-density-matrix-3}).
The reason why we do not choose equation~(\ref{general-2-qubit-density-matrix-3})
as the expression of the general density matrix
is as follows.
$\rho_{\mbox{\scriptsize AB}}$ defined in equation~(\ref{general-2-qubit-density-matrix-3})
does not include zero elements,
so that it is very difficult to calculate
a determinant of $\rho_{\mbox{\scriptsize AB}}^{\mbox{\scriptsize PT}}$
and eigenvalues of $\rho_{\mbox{\scriptsize AB}}\tilde{\rho}_{\mbox{\scriptsize AB}}$
explicitly.
By contrast,
if we choose equation~(\ref{general-2-qubit-density-matrix-2})
as the expression of $\rho_{\mbox{\scriptsize AB}}$,
it includes some zero elements in the matrix form
and
explicit calculations of determinants and eigenvalues are not so difficult.
Thus, although it has one extra real parameter,
we choose equation~(\ref{general-2-qubit-density-matrix-2}).

Now,
we construct $\rho_{\mbox{\scriptsize AB}}\tilde{\rho}_{\mbox{\scriptsize AB}}$
from $\rho_{\mbox{\scriptsize AB}}$ given by equation~(\ref{general-2-qubit-density-matrix-2}),
and calculate its eigenvalues
$\{\lambda_{i}:i\in\{1,2,3,4\}\}$.
The eigenvalues are solutions of the following quartic equation,
\begin{equation}
\det(
\rho_{\mbox{\scriptsize AB}}\tilde{\rho}_{\mbox{\scriptsize AB}}
-\lambda I
)
=0.
\label{equation-eigenvalues-0}
\end{equation}
Writing down equation~(\ref{equation-eigenvalues-0})
as a polynomial in $\lambda$,
we obtain
\begin{equation}
\lambda^{4}
+
f_{1}(\{r\})\lambda^{3}
+
f_{2}(\{r\})\lambda^{2}
+
f_{3}(\{r\})\lambda
+
f_{4}(\{r\})
=0,
\label{equation-eigenvalues-1}
\end{equation}
where
$f_{i}(\{r\})$ for $i\in\{1,2,3,4\}$ is a short form of
$f_{i}(r,s,t,u,v,w,q,\tau_{1},\tau_{2},\tau_{3})$.
An explicit form of $f_{1}(\{r\})$ is given by
\begin{equation}
f_{1}(\{r\})=-2[r(1-r-s-t)+st+u^{2}+v^{2}].
\label{explicit-form-f1}
\end{equation}
It is obvious that $f_{1}(\{r\})\leq 0$.
Because explicit forms of $f_{2}(\{r\})$, $f_{3}(\{r\})$ and $f_{4}(\{r\})$
are too complicated,
we give them in appendix~\ref{section-appendix}.

Next, we use the Ferrari's method for solving the quartic
equation~(\ref{equation-eigenvalues-1})
\cite{vanderWaerden1964, Birkhoff1950}.
We introduce a new variable $x$ as
\begin{equation}
x=\lambda-\frac{\Delta}{4},
\end{equation}
where
\begin{equation}
\Delta=-f_{1}(\{r\})(\geq 0).
\label{definition-Delta-0}
\end{equation}
Then, we rewrite the quartic equation
given by equation~(\ref{equation-eigenvalues-1})
as follows:
\begin{equation}
x^{4}+a(\{r\})x^{2}+b(\{r\})x+c(\{r\})=0.
\label{equation-eigenvalues-2}
\end{equation}
We pay attention to the fact that equation~(\ref{equation-eigenvalues-2})
does not include the third-order term $x^{3}$.
We give explicit forms of $a(\{r\})$, $b(\{r\})$ and $c(\{r\})$
in appendix~\ref{section-appendix}.

If we let $\{x_{i}:i\in\{1,2,3,4\}\}$
be solutions of equation~(\ref{equation-eigenvalues-2}),
we obtain the following relations:
\begin{eqnarray}
&&
\sum_{i}x_{i}=0, \nonumber \\
&&
\sum_{i<j}x_{i}x_{j}=a(\{r\}), \nonumber \\
&&
\sum_{i<j<k}x_{i}x_{j}x_{k}=-b(\{r\}), \nonumber \\
&&
x_{1}x_{2}x_{3}x_{4}=c(\{r\}).
\label{equation-eigenvalues-3}
\end{eqnarray}
The solutions of equation~(\ref{equation-eigenvalues-2})
$\{x_{i}:i\in\{1,2,3,4\}\}$
are given by
\begin{eqnarray}
x_{1}
&=&
P-\frac{1}{2}\sqrt{-\frac{b}{P}+Q}, \nonumber \\
x_{2}
&=&
P+\frac{1}{2}\sqrt{-\frac{b}{P}+Q}, \nonumber \\
x_{3}
&=&
-P-\frac{1}{2}\sqrt{\frac{b}{P}+Q}, \nonumber \\
x_{4}
&=&
-P+\frac{1}{2}\sqrt{\frac{b}{P}+Q},
\label{solution-x-0}
\end{eqnarray}
where $P$, $Q$ and $b$ are short forms of $P(\{r\})$, $Q(\{r\})$ and $b(\{r\})$.
Moreover, $P(\{r\})$ and $Q(\{r\})$ are given by
\begin{eqnarray}
P(\{r\})
&=&
\frac{1}{2\sqrt{6}}
\Bigl[
-4a(\{r\})
+\frac{2\sqrt[3]{2}R(\{r\})}{S^{1/3}(\{r\})}
+\sqrt[3]{4}S^{1/3}(\{r\})
\Bigr]^{1/2}, \nonumber \\
Q(\{r\})
&=&
\frac{1}{3}
\Bigl[
-4a(\{r\})
-\frac{\sqrt[3]{2}R(\{r\})}{S^{1/3}(\{r\})}
-\frac{1}{\sqrt[3]{2}}S^{1/3}(\{r\})
\Bigr], \nonumber \\
R(\{r\})
&=&
a^{2}(\{r\})+12 c(\{r\}), \nonumber \\
S(\{r\})
&=&
T(\{r\})
+
\sqrt{
-4R^{3}(\{r\})+T^{2}(\{r\})
}, \nonumber \\
T(\{r\})
&=&
2a^{3}(\{r\})
+27b^{2}(\{r\})
-72a(\{r\})c(\{r\}).
\label{solution-x-1}
\end{eqnarray}

Here, we remember
\begin{equation}
\lambda_{i}=x_{i}+\frac{\Delta}{4}
\quad
\mbox{for $i\in\{1,2,3,4\}$},
\end{equation}
where we do not put $\{\lambda_{i}\}$
in decreasing order.
The relations
$\lambda_{i}\geq 0$ for $i\in\{1,2,3,4\}$
and
$\Delta\geq 0$
are always valid.
Thus, $x_{i}$ for $i\in\{1,2,3,4\}$ has to be real.
Hence, we find that $P$ is real and $[\pm(b/P)+Q]$ are equal to or larger than zero.

From the above relations,
we obtain
\begin{equation}
x_{2}\geq x_{1},
\quad
x_{4}\geq x_{3}.
\label{relations-x}
\end{equation}
Thus, the maximum number of $\{x_{i}\}$
is $x_{2}$ or $x_{4}$.
Hence, from now on,
we assume $x_{2}\geq x_{4}$.
We consider the case where $x_{2}<x_{4}$ later.

Assuming $x_{2}\geq x_{4}$,
we can describe
the concurrence of $\rho_{\mbox{\scriptsize AB}}$
defined in equation~(\ref{general-2-qubit-density-matrix-2})
as
\begin{equation}
C(\rho_{\mbox{\scriptsize AB}})
=
\sqrt{\lambda_{2}}-\sqrt{\lambda_{1}}-\sqrt{\lambda_{3}}-\sqrt{\lambda_{4}}.
\label{concurrence-general-0}
\end{equation}
Here, we think around the following relations,
which are valid because of equations~(\ref{solution-x-0})
and (\ref{relations-x}),
\begin{eqnarray}
\sqrt{\lambda_{2}}-\sqrt{\lambda_{1}}
&\geq&
0, \nonumber \\
\sqrt{\lambda_{3}}+\sqrt{\lambda_{4}}
&\geq&
0.
\end{eqnarray}
Thus, we can rewrite $C(\rho_{\mbox{\scriptsize AB}})$
given by equation~(\ref{concurrence-general-0})
as
\begin{eqnarray}
C(\rho_{\mbox{\scriptsize AB}})
&=&
\sqrt{(\sqrt{\lambda_{2}}-\sqrt{\lambda_{1}})^{2}}
-
\sqrt{(\sqrt{\lambda_{3}}+\sqrt{\lambda_{4}})^{2}} \nonumber \\
&=&
\sqrt{(\Delta/2)+x_{1}+x_{2}-2\sqrt{[(\Delta/4)+x_{1}][(\Delta/4)+x_{2}]}} \nonumber \\
&&
-
\sqrt{(\Delta/2)+x_{3}+x_{4}+2\sqrt{[(\Delta/4)+x_{3}][(\Delta/4)+x_{4}]}}.
\label{concurrence-general-1}
\end{eqnarray}

The necessary and sufficient condition for inseparability of $\rho_{\mbox{\scriptsize AB}}$
is given by $C(\rho_{\mbox{\scriptsize AB}})>0$.
Thus, from equation~(\ref{concurrence-general-1}),
the necessary and sufficient condition for inseparability of $\rho_{\mbox{\scriptsize AB}}$
can be rewritten as
\begin{eqnarray}
&&
x_{1}+x_{2}-2\sqrt{[(\Delta/4)+x_{1}][(\Delta/4)+x_{2}]} \nonumber \\
&&
\quad
>x_{3}+x_{4}+2\sqrt{[(\Delta/4)+x_{3}][(\Delta/4)+x_{4}]}.
\label{general-entangled-condition-0}
\end{eqnarray}
Moreover,
from equation~(\ref{solution-x-0}),
we rewrite equation~(\ref{general-entangled-condition-0})
as follows:
\begin{equation}
2P
>
\sqrt{[(\Delta/4)+x_{1}][(\Delta/4)+x_{2}]}
+
\sqrt{[(\Delta/4)+x_{3}][(\Delta/4)+x_{4}]}
(\geq 0).
\label{general-entangled-condition-1}
\end{equation}
[In the above derivation,
we use $\lambda_{i}=x_{i}+(\Delta/4)\geq 0\forall i$.]
Because both the right-hand and the left-hand sides of
inequality~(\ref{general-entangled-condition-1})
are equal to or larger than zero,
we can square both sides of inequality~(\ref{general-entangled-condition-1}) respectively,
and we obtain
\begin{eqnarray}
4P^{2}
&>&
[(\Delta/4)+x_{1}][(\Delta/4)+x_{2}]
+
[(\Delta/4)+x_{3}][(\Delta/4)+x_{4}] \nonumber \\
&&
\quad
+
2\sqrt{[(\Delta/4)+x_{1}][(\Delta/4)+x_{2}][(\Delta/4)+x_{3}][(\Delta/4)+x_{4}]}.
\label{general-entangled-condition-2}
\end{eqnarray}

Looking at equations~(\ref{equation-eigenvalues-3})
and (\ref{solution-x-0}),
we notice that we can rewrite inequality~(\ref{general-entangled-condition-2})
as
\begin{equation}
2P^{2}
>
\frac{\Delta^{2}}{8}
-\frac{Q}{2}
+2
\sqrt{(\Delta/4)^{4}+a(\Delta/4)^{2}-b(\Delta/4)+c}.
\label{general-entangled-condition-3}
\end{equation}
Substituting equation~(\ref{solution-x-1}) into inequality~(\ref{general-entangled-condition-3}),
we obtain
\begin{equation}
-a-\frac{\Delta^{2}}{8}-2\sqrt{(\Delta/4)^{4}+a(\Delta/4)^{2}-b(\Delta/4)+c}>0.
\label{general-entangled-condition-4}
\end{equation}

Here, we pay attention to the following relation,
which is obtained from\\
equations~(\ref{determinant-positive-or-zero-0}),
(\ref{explicit-form-f1}), (\ref{definition-Delta-0}),
(\ref{explicit-forms-f234}) and (\ref{explicit-forms-abc}),
\begin{equation}
(\Delta/4)^{4}+a(\Delta/4)^{2}-b(\Delta/4)+c
=
[\det(\rho_{\mbox{\scriptsize AB}})]^{2}.
\end{equation}
Thus, from equation~(\ref{determinant-positive-or-zero-0}),
we obtain
\begin{equation}
\sqrt{(\Delta/4)^{4}+a(\Delta/4)^{2}-b(\Delta/4)+c}
=
\det(\rho_{\mbox{\scriptsize AB}})(\geq 0).
\label{relation-determinant-A}
\end{equation}

Then, using equation~(\ref{relation-determinant-A}),
we can rewrite equation~(\ref{general-entangled-condition-4})
as
\begin{eqnarray}
D(\{r\})
&=&
rsq^{2}+rtw^{2}+(st-u^{2})[v^{2}-r(1-r-s-t)] \nonumber \\
&&
-2ruwq\cos(\tau_{1}-\tau_{2}-\tau_{3}) \nonumber \\
&>&0.
\label{general-entangled-condition-5}
\end{eqnarray}
Hence, we find that inequality~(\ref{general-entangled-condition-5})
is the necessary and sufficient condition for
inseparability of $\rho_{\mbox{\scriptsize AB}}$
defined in equation~(\ref{general-2-qubit-density-matrix-2}).

In the discussions given above, we assume $x_{2}\geq x_{4}$.
If $x_{2}<x_{4}$,
we can give similar discussions
and we obtain
inequality~(\ref{general-entangled-condition-5})
as the necessary and sufficient condition for
inseparability of $\rho_{\mbox{\scriptsize AB}}$, as well.
Therefore,
we conclude that
the necessary and sufficient condition for
inseparability of $\rho_{\mbox{\scriptsize AB}}$
derived from Hill-Wootters formula
is given by
inequality~(\ref{general-entangled-condition-5}).

On the other hand,
according to the Peres-Horodecki criterion,
the necessary and sufficient condition for
inseparability of $\rho_{\mbox{\scriptsize AB}}$,
which is defined in equation~(\ref{general-2-qubit-density-matrix-2}),
is given by
\begin{equation}
\det(\rho_{\mbox{\scriptsize AB}}^{\mbox{\scriptsize PT}})<0.
\label{inseparability-general-Peres-Horodecki-0}
\end{equation}
Writing equation~(\ref{inseparability-general-Peres-Horodecki-0})
in an explicit form,
we obtain
\begin{equation}
\det(\rho_{\mbox{\scriptsize AB}}^{\mbox{\scriptsize PT}})
=-D(\{r\})
<0.
\label{inseparability-general-Peres-Horodecki-1}
\end{equation}
Inequality~(\ref{inseparability-general-Peres-Horodecki-1})
is equivalent to inequality~(\ref{general-entangled-condition-5}).

Therefore,
we succeed in deriving the Peres-Horodecki criterion for two-qubit mixed states
from the Hill-Wootters formula in a direct manner.

\section{Separability for a convex combination
of a separable pure state and an inseparable pure state}
\label{section-convex-combination-separable-inseparable}
In this section,
we consider an example of a convex combination of a separable pure state and an inseparable pure state,
\begin{equation}
\rho_{\mbox{\scriptsize AB}}
=
p
|\phi_{s}\rangle_{\mbox{\scriptsize AB}}
{}_{\mbox{\scriptsize AB}}\langle\phi_{s}|
+
(1-p)
|\psi_{e}\rangle_{\mbox{\scriptsize AB}}
{}_{\mbox{\scriptsize AB}}\langle\psi_{e}|,
\label{density-matrix-separable-pure-entangled-pure-0}
\end{equation}
where
$0\leq p\leq 1$,
$|\phi_{s}\rangle_{\mbox{\scriptsize AB}}$ is a normalized separable ket vector
and
$|\psi_{e}\rangle_{\mbox{\scriptsize AB}}$ is a normalized inseparable ket vector.
Under the assumption of equation~(\ref{density-matrix-separable-pure-entangled-pure-0}),
we can derive the Peres-Horodecki criterion
from
the Hill-Wootters formula
in a direct manner
without difficulty,
so that this can be a concrete example of the results obtained
in section~\ref{section-general-case-Hill-Wootters-Peres-Horodecki}.

At first, we think about a couple of SU(2) transformations
$U_{1,\mbox{\scriptsize A}}$ and $U_{2,\mbox{\scriptsize B}}$,
which cause the following transformation to
the separable pure state given in equation~(\ref{density-matrix-separable-pure-entangled-pure-0}),
\begin{eqnarray}
&&
|\phi_{s}\rangle_{\mbox{\scriptsize AB}}
=
|\phi_{1}\rangle_{\mbox{\scriptsize A}}
\otimes
|\phi_{2}\rangle_{\mbox{\scriptsize B}} \nonumber \\
&\rightarrow &
(U_{1,\mbox{\scriptsize A}}\otimes U_{2,\mbox{\scriptsize B}})
|\phi_{s}\rangle_{\mbox{\scriptsize AB}}
=
U_{1,\mbox{\scriptsize A}}|\phi_{1}\rangle_{\mbox{\scriptsize A}}
\otimes
U_{2,\mbox{\scriptsize B}}|\phi_{2}\rangle_{\mbox{\scriptsize B}} \nonumber \\
&&
=
|0\rangle_{\mbox{\scriptsize A}}
\otimes
|0\rangle_{\mbox{\scriptsize B}}.
\label{diagonalization_system_AB}
\end{eqnarray}
The couple of the unitary transformations $U_{1,\mbox{\scriptsize A}}$ and $U_{2,\mbox{\scriptsize B}}$
that satisfy equation~(\ref{diagonalization_system_AB})
always exists
and we can choose $U_{1,\mbox{\scriptsize A}}$ and $U_{2,\mbox{\scriptsize B}}$
independently with each other.
Moreover, applying $U_{1,\mbox{\scriptsize A}}\otimes U_{2,\mbox{\scriptsize B}}$
to the system AB neither increases nor decreases the entanglement between the qubits A and B,
because it is a local operation.
Thus, applying $U_{1,\mbox{\scriptsize A}}\otimes U_{2,\mbox{\scriptsize B}}$
never gives actual effects on both the Peres-Horodecki criterion
and the Hill-Wootters formula.

From the above discussions,
we can use the following density matrix as a general form
instead of the density matrix given by equation~(\ref{density-matrix-separable-pure-entangled-pure-0}),
\begin{equation}
\rho_{\mbox{\scriptsize AB}}
=
p
\left(
\begin{array}{cccc}
1 & 0 & 0 & 0 \\
0 & 0 & 0 & 0 \\
0 & 0 & 0 & 0 \\
0 & 0 & 0 & 0
\end{array}
\right)
+
(1-p)
|\psi_{e}\rangle_{\mbox{\scriptsize AB}}
{}_{\mbox{\scriptsize AB}}\langle\psi_{e}|,
\label{density-matrix-separable-pure-entangled-pure-1}
\end{equation}
where
$0\leq p \leq 1$,
\begin{equation}
|\psi_{e}\rangle_{\mbox{\scriptsize AB}}
=
\left(
\begin{array}{c}
a \\
b \exp(i\theta_{1}) \\
c \exp(i\theta_{2}) \\
\sqrt{1-a^{2}-b^{2}-c^{2}} \exp(i\theta_{3})
\end{array}
\right),
\end{equation}
and
$a\geq 0$, $b\geq 0$, $c\geq 0$, $1-a^{2}-b^{2}-c^{2}\geq 0$,
$0 \leq \theta_{i}<2\pi$ for $i\in\{1,2,3\}$.
Moreover,
because $|\psi_{e}\rangle_{\mbox{\scriptsize AB}}$ is an entangled state,
we assume its concurrence to be positive as follows:
\begin{eqnarray}
C(|\psi_{e}\rangle_{\mbox{\scriptsize AB}})
&=&
2
[
a^{2}(1-a^{2}-b^{2}-c^{2})+b^{2}c^{2} \nonumber \\
&&\quad
-2abc\sqrt{1-a^{2}-b^{2}-c^{2}}\cos(\theta_{1}+\theta_{2}-\theta_{3})
]^{1/2} \nonumber \\
&>&0.
\end{eqnarray}

From the density matrix given by equation~(\ref{density-matrix-separable-pure-entangled-pure-1}),
we calculate eigenvalues of $\rho_{\mbox{\scriptsize AB}}\tilde{\rho}_{\mbox{\scriptsize AB}}$.
We write them as $\lambda_{1}\geq\lambda_{2}\geq\lambda_{3}\geq\lambda_{4}\geq 0$,
where
\begin{equation}
\lambda_{1}=X+\frac{1}{2}\sqrt{Y},
\quad
\lambda_{1}=X-\frac{1}{2}\sqrt{Y},
\quad
\lambda_{3}=\lambda_{4}=0,
\end{equation}
\begin{eqnarray}
X
&=&
\frac{1}{2}(1-p)
[(1-p)C^{2}(|\psi_{e}\rangle_{\mbox{\scriptsize AB}})
+2p(1-a^{2}-b^{2}-c^{2})], \nonumber \\
Y
&=&
(1-p)^{3}C^{2}(|\psi_{e}\rangle_{\mbox{\scriptsize AB}})
[(1-p)C^{2}(|\psi_{e}\rangle_{\mbox{\scriptsize AB}})
+4p(1-a^{2}-b^{2}-c^{2})]. \nonumber \\
\end{eqnarray}
From the above equations,
we can obtain
$X\geq 0$ and $Y\geq 0$ at ease.

Thus, we can write the concurrence of $\rho_{\mbox{\scriptsize AB}}$
as
\begin{equation}
C(\rho_{\mbox{\scriptsize AB}})=\sqrt{\lambda_{1}}-\sqrt{\lambda_{2}}.
\end{equation}
Hence, we find that
the necessary and sufficient condition of $C(\rho_{\mbox{\scriptsize AB}})>0$
($\rho_{\mbox{\scriptsize AB}}$ is inseparable)
is $Y>0$.

On the other hand,
defining the density matrix $\rho_{\mbox{\scriptsize AB}}$
as equation~(\ref{density-matrix-separable-pure-entangled-pure-1}),
we can obtain the determinant of $\rho_{\mbox{\scriptsize AB}}^{\mbox{\scriptsize PT}}$
in the form,
\begin{equation}
\det(\rho_{\mbox{\scriptsize AB}}^{\mbox{\scriptsize PT}})
=
-\frac{Y}{16}.
\end{equation}

Thus,
the determinant of $\rho_{\mbox{\scriptsize AB}}^{\mbox{\scriptsize PT}}$
is negative
if and only if $\rho_{\mbox{\scriptsize AB}}$ is inseparable,
where we assume that $\rho_{\mbox{\scriptsize AB}}$ is given
by equation~(\ref{density-matrix-separable-pure-entangled-pure-0}).
Hence,
we derive the Peres-Horodecki criterion
from the Hill-Wootters formula in a direct manner
on condition that the density matrix is given by
a convex combination
of a separable pure state and an inseparable pure state.

\section{Discussions}\label{section-discussions}
In this paper,
we investigate connections between
the Peres-Horodecki criterion for the two-qubit states
and
the Hill-Wootters formula for the entanglement of formation.
In this study,
the following expression being equivalent to the Peres-Horodecki criterion
plays an important role:
the two-qubit mixed state is inseparable
if and only if the determinant of the partial transpose of its density matrix
is less than zero.
In reference~\cite{Augusiak2008},
Augusiak {\it et al}. show that an entanglement measure for two-qubit state can be
constructed from $\max\{0,-\det(\rho_{\mbox{\scriptsize AB}}^{\mbox{\scriptsize PT}})\}$.
The authors suppose
that there are an infinite number of entanglement measures for two-qubit systems.

\section*{Acknowledgements}
While H. A. was working as Postdoctoral Research Assistant
at Clarendon Laboratory in 2000,
Patrick Hayden told him that
the partially transposed density matrix of the inseparable two-qubit mixed state
has one negative eigenvalue at most
and this fact can be derived from the so-called Lewenstein-Sanpera decomposition
\cite{Lewenstein1998, Hayden2000}.
This personal communication inspires the authors
to study main themes of this paper.
H. A. also thanks the colleagues of IMS Lab. for encouragement.

\appendix
\section{Appendix}\label{section-appendix}
Explicit forms of $f_{i}(\{r\})$ for $i\in\{2,3,4\}$ defined in equation~(\ref{equation-eigenvalues-1})
are given by
\begin{eqnarray}
f_{2}(\{r\})
&=&
-r^{4}-2r^{3}(1-r-s-t)+s^{2}t^{2}+2u^{2}v^{2}
+2st(2u^{2}-v^{2}) \nonumber \\
&&
+(u^{2}+v^{2})^{2}
+r^{2}[1+(s-t)^{2}-2(s+t)+2(u^{2}-2v^{2})] \nonumber \\
&&
-2r[q^{2}s-s(u^{2}-2v^{2})+(1-t)(u^{2}-2v^{2}-2st) \nonumber \\
&&\quad\quad
+t(2s^{2}+w^{2})] \nonumber \\
&&
+4qrw[2u\cos(\tau_{1}-\tau_{2}-\tau_{3})-v\cos(\tau_{2}-\tau_{3})], \nonumber \\
f_{3}(\{r\})
&=&
2
\Biggl[
-stu^{4}
-v^{4}[u^{2}+r\eta(\{r\})]
+rs[st+r\eta(\{r\})][q^{2}-t\eta(\{r\})] \nonumber \\
&&\quad
-su^{2}(rq^{2}+t[st-2r\eta(\{r\})])
-rw^{2}(2rq^{2}+t[u^{2}+v^{2}-st-r\eta(\{r\})]) \nonumber \\
&&\quad
-v^{2}(rsq^{2}+[u^{2}-r\eta(\{r\})]^{2}-2st[u^{2}+r\eta(\{r\})]) \nonumber \\
&&\quad
+4ru
\Bigl(
tvw^{2}\cos(\tau_{1}-2\tau_{2})
+q[svq\cos(\tau_{1}-2\tau_{3}) \nonumber \\
&&\quad\quad
-(st+v^{2})w\cos(\tau_{1}-\tau_{2}-\tau_{3})]
\Bigr) \nonumber \\
&&\quad
-2rvwq[st+u^{2}-v^{2}-r\eta(\{r\})]\cos(\tau_{2}-\tau_{3})
\Biggr] \nonumber \\
&&\quad\quad\quad
\mbox{where $\eta(\{r\})=1-r-s-t$,}\nonumber \\
f_{4}(\{r\})
&=&
\Biggl[
rsq^{2}
+[u^{2}-r(1-r-s-t)](st-v^{2})+rtw^{2} \nonumber \\
&&\quad
-2rvwq\cos(\tau_{2}-\tau_{3})
\Biggr]^{2}.
\label{explicit-forms-f234}
\end{eqnarray}

The explicit forms of $a(\{r\})$, $b(\{r\})$ and $c(\{r\})$
defined in equation~(\ref{equation-eigenvalues-2}) are given by
\begin{eqnarray}
a(\{r\})
&=&
f_{2}(\{r\})-(3/8)\Delta^{2}, \nonumber \\
b(\{r\})
&=&
f_{3}(\{r\})-(1/8)\Delta[\Delta^{2}-4f_{2}(\{r\})], \nonumber \\
c(\{r\})
&=&
f_{4}(\{r\})-(1/256)\Delta[3\Delta^{3}-16\Delta f_{2}(\{r\})-64f_{3}(\{r\})].
\label{explicit-forms-abc}
\end{eqnarray}

\end{document}